\UseRawInputEncoding
\documentclass[sigconf]{acmart}

\usepackage{tcolorbox}
\usepackage[ruled,vlined,linesnumbered]{algorithm2e}
\usepackage{graphicx}
\usepackage{float}
\usepackage{booktabs}
\usepackage{multirow}
\usepackage{color}
\usepackage{xcolor}

\newtcbox{\mybox}[1][red]
  {on line, arc = 0pt, outer arc = 0pt,
    colback = #1!10!white, colframe = #1!50!black,
    boxsep = 0pt, left = 1pt, right = 1pt, top = 2pt, bottom = 2pt,
    boxrule = 0pt, bottomrule = 1pt, toprule = 1pt}
\AtBeginDocument{%
  }

\setcopyright{acmlicensed}
\copyrightyear{2024}
\acmYear{2024}
\acmDOI{XXXXXXX.XXXXXXX}

\renewcommand\footnotetextcopyrightpermission[1]{} 
\acmConference[ASE'24]{The 39th IEEE/ACM International Conference on Automated Software Engineering}{October 27 -- November 01,
  2024}{Sacramento, CA, USA}
\acmISBN{978-1-4503-XXXX-X/18/06}

\begin{document}

\title{DNLSAT: A Dynamic Variable Ordering MCSAT Framework for Nonlinear Real Arithmetic}


\author{Zhonghan Wang}
\affiliation{
    \institution{State Key Laboratory of Computer Science, Institute of Software, Chinese Academy of Sciences\\
School of Computer Science and Technology, University of Chinese Academy of Sciences}
    \city{Beijing}
    \country{China}
}
\email{wangzh@ios.ac.cn}
\renewcommand{\shortauthors}{Zhonghan Wang}

\begin{abstract}
  Satisfiability modulo nonlinear real arithmetic theory (SMT(NRA)) solving is essential to multiple applications, including program verification, program synthesis and software testing. In this context, recently model constructing satisfiability calculus (MCSAT) has been invented to directly search for models in the theory space. Although following papers discussed practical directions and updates on MCSAT, less attention has been paid to the detailed implementation. In this paper, we present an efficient implementation of dynamic variable orderings of MCSAT, called dnlsat. We show carefully designed data structures and promising mechanisms, such as branching heuristic, restart, and lemma management. Besides, we also give a theoretical study of potential influences brought by the dynamic variablr ordering. The experimental evaluation shows that dnlsat accelerates the solving speed and solves more satisfiable instances than other state-of-the-art SMT solvers.\\
  \textbf{Demonstration Video:} \url{https://youtu.be/T2Z0gZQjnPw}. \\
  \textbf{Code:} \url{https://github.com/yogurt-shadow/dnlsat/tree/master/code} \\
  \textbf{Benchmark:} \url{https://zenodo.org/records/10607722/files/QF_NRA.tar.zst?download=1}
\end{abstract}






\maketitle

\section{Introduction}
Satisfiability Modulo Theories (SMT) extends boolean satisfiability (SAT) problems into different background theories, such as linear and nonlinear arithmetic, uninterpretted functions, strings and arrays~\cite{BarrettT18}. Among them, nonlinear real arithmetic (NRA), represents logical formulas with polynomial constraints, and enables variables to be real numbers. Nonlinear real arithmetic solving is essential to various domains in computer science, including both academical and industrial applications. For example, since nonlinear arithmetic is good at representing differential equations along the continuous systems, SMT(NRA) solvers are efficient tools to predict and control behavious in cyber physical systems~\cite{CPS1, CPS2, CPS3}. Other applications including ranking function generation~\cite{LeikeH15, HeizmannHLP13} used for termination analysis and nonlinear hybrid automata~\cite{CimattiMT12} analysis also take advantage of SMT(NRA) solvers. Recently, SMT(NRA) solvers also helps a lot in artificial intelligence related studies, like the verification and repairing of neural networks~\cite{NN1, NN2, NN3, NN4}. In conclusion, designing powerful SMT(NRA) solvers is essential and fundamental for different research directions. 

The traditional method to deal with nonlinear arithmetic is CDCL(T), with a theory solver as a black box used to check the consistence of theory literals and return a new lemma. Recently, MCSAT framework has been introduced to enroll the theory solver into the search space, and lead the literal level search into variable level. However, although MCSAT is efficient in solving most instances, many ideas that have been proved to be effective in SAT solving are not applied. In this paper, we introduce a new solver named dnlsat, which brings usually used systematic search heuristics into nonlinear arithmetic solving. Our experimental results demonstrate that dnlsat is competitive on satisfiable problems, which means that it might be a powerful tool used in program termination analysis and bug finding.

\section{Related Work}
model constructing satisfiability calculus (MCSAT) has been widely applied to solve SMT problems over different theories. The spirit of MCSAT is to directly assign arithmetic variables with a theory solver incorporated in the solving process, rather than assign literals like CDCL(T) does. For nonlinear real arithmetic, NLSAT is an efficient implementation of MCSAT, which uses cylindrical algebraic decomposition (CAD) for explanation when encountering conflicts. The seaech process is repeated until a solution is found or it is determined that the problem is unsatisfiable.

Thanks to the MCSAT framework, many ideas have been applied to tackle directly with arithmetic variables. For example, some studies have tried to investigate the branching heuristic of MCSAT, and talk about the related completeness problem~\cite{MCSATOrder}. Other work has been discussing the proof complexity of MCSAT algorithm~\cite{kremer2021proof}. In conclusion, it is interesting to bring heuristics used in SAT solving to the MCSAT framework, with a transfer from boolean space to the real space.

\section{Architecture}
\begin{figure}
    \centering
    \includegraphics[width=0.45\textwidth]{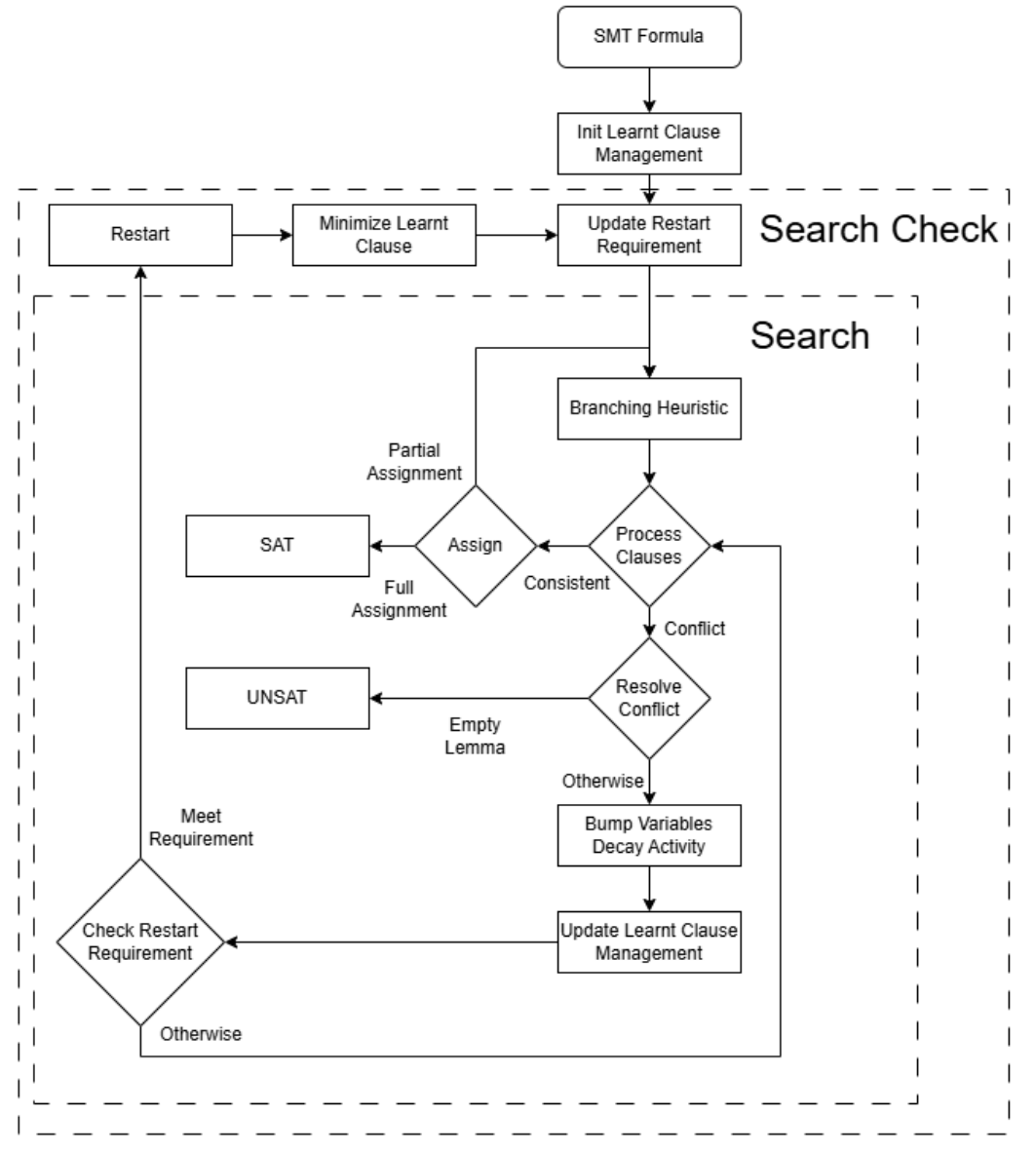}
    \caption{Overall Structure of dnlsat.}
    \label{fig:structure}
\end{figure}
The general architecture of dnlsat is shown in Fig. ~\ref{fig:structure}. Most of the parts are borrowed from CDCL SAT solvers, such as minisat~\cite{Srensson2005MiniSatV}. We first give our implementation of dynamic variable orderings (i.e. branching heuristic), as suggested in ~\cite{MCSATOrder}. As proposed in ~\cite{MCSATOrder, MCSAT2}, we talk about the implementation of detecting a univariate clause, with a consideration of root atoms. Second, some kinds of MCSAT trails like level and stage are also discussed to better manage conflict analysis and backtrack. Third, we give a theoretical analysis of CAD projection orders. Since different projection orders will generate different forms of lemmas, we talk about the set of orders that can be used to resolve conflicts. Fourth, we implement a lemma management mechanism to periodically delete useless lemmas, which is proved to be very powerful when encountering disabled root atoms. Finally, we introduce restart mechanism into systematic search of SMT solvers.
\subsection{Branching Heuristic}
We implement the following variable orderings as suggested in ~\cite{MCSATOrder}.
\begin{itemize}
    \item \textbf{Default.} This is the default setting in NLSAT. Boolean variables are decided before any theory decision. Arithmetic variables are ordered by their maximum degrees in polynomial constraints.
    \item \textbf{Boolean-VSIDS.} Boolean variables are always decided before arithmetic variables. Two variables with the same type are ordered by their activity.
    \item \textbf{Theory-VSIDS.} Arithmetic variables are always decided before boolean variables. Two variables with the same type are ordered by their activity.
    \item \textbf{Uniform-VSIDS.} Whatever type the variables are, they are ordered only by their activity.
\end{itemize}

\subsection{Projection Order}
Projection order is essential for CAD algorithm. However, in real applications like SMT solvers, the order is actually not random when considering the power of generated lemmas. Given a polynomial set $ps$, and a projection order $\{v_1, v_2, ..., v_k\}$, each time the projection method eliminates a variable and generates a root atom according to that variable. In this case, root atoms are generated in the following form:
\begin{align*}
    v_1 \quad \sim & \quad root(p_1(v_1, v_2, v_3, ..., v_k), i_1) \\
    v_2 \quad \sim & \quad root(p_2(v_3, v_4, ..., v_k), i_2) \\
    ... \quad \sim & \quad ...\\
    v_k \quad \sim & \quad root(p_k(v_k), i_k)
\end{align*}
where polynomial $p_k$ is univariate to $v_k$. To deal with the problem of useless root atoms as discuss in ~\cite{MCSATOrder}, the projection order is the reverse of the order of assigned variables.

\subsection{Lemma Management}
We borrow the idea of periodically forgetting useless lemmas from minisat~\cite{Srensson2005MiniSatV}.
\subsubsection{Requirement}
We try to minimize the learnt clauses only when we jump out the search process and restart. We record a counter named \emph{learntsize\_adjust\_cnt}. Only when \emph{learntsize\_adjust\_cnt} counts down to zero, the minimize procedure takes effect. The increasement factor of \emph{learntsize\_adjust\_cnt} is named \emph{learntsize\_adjust\_inc}, which tries to improve the threshold of minimize process alongwith the systematic search.

\subsubsection{Clause Activity}
Similar to the activity used in VSIDS, the activity of learnt clauses also record their status of being involved in conflict analysis. Whenever we find a learnt clause in the resolve process, its activity is bumped by \emph{clause\_bump}. To inherit the spirit that focus more on recent conflicts, \emph{clause\_bump} is increased by a factor \emph{clause\_decay} after each update.

\subsubsection{Minimize Process}
To better preserve useful lemmas, we set a parameter named \emph{max\_learnts}. Only when the current database contains a number of learnt clauses larger than \emph{max\_learnts}, our minimize process is executed. We sort the learnt clauses by their activities and try to delete those most inactive clauses by a half. Short clauses that contain less than three literals are preserved.

\subsection{Restart}
\subsubsection{Requirement}
We check the requirement of restart after each iteration of search process. When the number of conflicts surpass a threshold, the restart process executes. In our implementation, the threshold conflict times of enabling restarts is an exponential sequential calculated as $threshold = restart\_first * restart\_base^{restart\_times}$.

\subsubsection{Information Preserve}
Most information about the search process is preserved in our restart mechanism, including the activity of variables and clauses, the number of total decisions and a part of learnt clauses.

\section{Implementation and Usage}
Our algorithm is implemented on top of the nlsat solver in Z3, based on the existing library for polynomials and algebraic numbers. The source code and other resources of dnlsat is availabe at: \url{https://github.com/yogurt-shadow/dnlsat}. We present hyperparameters used in our implementation in Table ~\ref{tab:parameters}.

\textbf{Usage:} The compilation and execution method are the same with Z3 solver. Dnlsat accepts an input with SMT-LIB v2.6 format~\footnote{https://smtlib.github.io/jSMTLIB/SMTLIBTutorial.pdf} (\texttt{example.smt2}). We assume the user enters the \texttt{code} folder. To compile the overall project, simply run python command

\begin{center}
\texttt{python scripts/mk\_make.py}\\
\texttt{cd build}\\
\texttt{make -j<job\_number>}
\end{center}

To solve a SMT instance, simply run
\begin{center}
\texttt{./z3 example.smt2}
\end{center}

\begin{table}[]
\small
    \centering
    \begin{tabular}{c|c|c}
      Symbol & Description & ~Value~ \\ \hline
      $\mathit{arith\_decay}$ & Decay\_factor for arithmetic variables in VSIDS & 0.95 \\
      $\mathit{bool\_decay}$ & Decay\_factor for boolean variables in VSIDS & 0.95 \\
      $\mathit{arith\_bump}$ & Incremental amount of arithmetic activity & 1 \\
      $\mathit{bool\_bump}$ & Incremental amount of boolean activity & 1 \\
      $\mathit{lemma\_conf}$ & Initial conflict times for deleting lemmas & 100 \\
    $\mathit{lemma\_conf\_inc}$ & Incremental factor of conflict for lemmas & 1.5 \\ 
    $\mathit{learntsize\_factor}$ & The ratio of preserved lemma of origin clauses & $\frac{1}{3}$ \\
    $\mathit{clause\_decay}$ & The decay factor of clause's activity & 0.999 \\
    $\mathit{restart\_first}$ & Conflict times of the first restart & 100 \\
    $\mathit{restart\_inc}$ & Incremental factor of conflict for restart & 1.5\\
      $\mathit{clause\_bump}$ & Incremental amount of clause activity & 1 \\
    \hline
    \end{tabular}
    \vspace{2mm}
    \caption{Hyperparameters of dnlsat}
    \label{tab:parameters}
\end{table}

\section{Evaluation}
In this section, we demonstrate the performance of dnlsat on SMT(NRA) benchmarks. We first describe our experimental environment, benchmarks and baseline solvers. Then we compare different versions of branching heuristics. Finally, we present the comparison between dnlsat and other state-of-the-art solvers in terms of solved satisfiable and unsatisfiable instances.

\subsection{Experiment Preliminaries}
\subsubsection{Environment} We evaluate our experiments on a server with Intel Xeon Platinum 8153 processor at 2.00 GHz. The time limit of running time for each instance is 1200 seconds.
\subsubsection{Benchmarks} We choose the SMT-LIB~\cite{BarFT-SMTLIB} benchmark repository QF\_NRA track\footnote{https://zenodo.org/records/10607722} for our evaluation, which contains 12134 instances in total.
\subsubsection{Baseline Solvers} We compare our solver with three most powerful solvers in SMT-COMP, including Z3 (version 4.13.1)~\cite{MouraB08}, CVC5 (version 1.0.2)~\cite{BarbosaBBKLMMMN22} and YICES2 (version 2.6.2)~\cite{Dutertre14}. Note that we do not make any modification to their source code, and preserve all the different algorithms in the portfolio solvers.

\subsection{Comparison between different branching heuristics}
\begin{figure}
     \centering
     \includegraphics[width=0.45\textwidth]{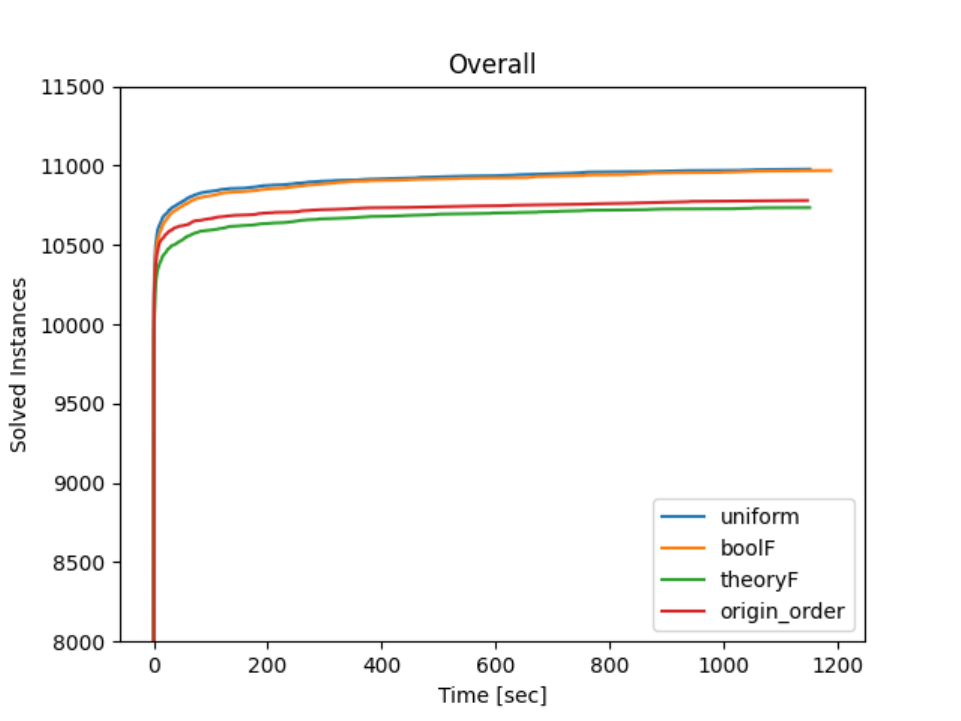}
    \caption{The number of solved instances by different branching heuristics within distinct time limits.}
\label{fig:bh}
\end{figure}

Figure ~\ref{fig:bh} shows the number of solved instances by different branching heuristics within distinct time limits. It is found that uniform-vsids and bool-vsids solve more instances than theory-vsids and the default static ordering. Our results are different from previous studies~\cite{MCSATOrder}, the main difference is that SMT-RAT implements multiple pulgins used for explanation, such as virtual substitution method (VS), Fourier-Motzkin variable elimination method (FM) and cylindrical algebraic decomposition method (CAD), while we only preserve CAD plugin for backend explanation part. Due to our efficient data structures and other heuristics, we solve more instances than SMT-RAT as described in ~\cite{MCSATOrder}.

\subsection{Comparison with SOTA solvers}
Table ~\ref{tab:instances} shows the overall solved instances of dnlsat and other state-of-the-art solvers, which are divided into different categories. It is noticed that dnlsat is competitive with other portfolio solvers with only MCSAT algorithm implemented. Specifically, dnlsat solves the most satisfiable instances. Although dnlsat only solved third most overall instances, the difference between the second solver YICES2 is only 7 instances. 


\begin{table}[]
\centering
\tiny
\begin{tabular}{|*{15}{c|}}
\hline
Category & \multicolumn{2}{|c|}{\#inst} & Z3 & YICES2 & CVC5 & dnlsat \\\hline
\multirow{2}*[-1.5ex]{20161105-Sturm-MBO} & \multirow{2}*[-1.5ex]{405} 
& SAT & 0 & 0 & 0 & 0 \\\cline{3-7}
& & UNSAT & 124 & \textbf{285} & \textbf{285} &  39 \\\cline{3-7}
& & SOLVED & 124 & \textbf{285} & \textbf{285} & 39 \\\hline
\multirow{2}*[-1.5ex]{20161105-Sturm-MGC} & \multirow{2}*[-1.5ex]{9} 
& SAT & \textbf{2} & 0 & 0 &  \textbf{2} \\\cline{3-7}
& & UNSAT & \textbf{7} & 0 & 0 &  6 \\\cline{3-7}
& & SOLVED & \textbf{9} & 0 & 0 & 8 \\\hline
\multirow{2}*[-1.5ex]{20170501-Heizmann} & \multirow{2}*[-1.5ex]{69} 
     & SAT & \textbf{2} & 0 & 1 & \textbf{2} \\\cline{3-7}
 & & UNSAT & 1 & 12 & 9 &  \textbf{19} \\\cline{3-7}
& & SOLVED & 3 & 12 & 10 & \textbf{21} \\\hline
\multirow{2}*[-1.5ex]{20180501-Economics-Mulligan} & \multirow{2}*[-1.5ex]{135} 
     & SAT & \textbf{93} & 91 & 89 & 92 \\\cline{3-7}
 & & UNSAT & 39 & 39 & 35 & \textbf{41} \\\cline{3-7}
& & SOLVED & 132 & 130 & \textbf{134} & 131 \\\hline
\multirow{2}*[-1.5ex]{2019-ezsmt} & \multirow{2}*[-1.5ex]{63} 
     & SAT & \textbf{56} & 52 & 50 &  29 \\\cline{3-7}
 & & UNSAT & \textbf{2} & \textbf{2}& \textbf{2} & \textbf{2} \\\cline{3-7}
& & SOLVED & \textbf{58} & 54 & 52 & 31 \\\hline
\multirow{2}*[-1.5ex]{20200911-Pine} & \multirow{2}*[-1.5ex]{245} 
     & SAT & 234 & \textbf{235} & 199 & 234 \\\cline{3-7}
 & & UNSAT & 6 & \textbf{8} & 5 & 5 \\\cline{3-7}
& & SOLVED & 240 & \textbf{243} & 204 &  239 \\\hline
\multirow{2}*[-1.5ex]{20211101-Geogebra} & \multirow{2}*[-1.5ex]{112} 
     & SAT & \textbf{110} & 99 & 91 & 100 \\\cline{3-7}
 & & UNSAT & 0 & 0 & 0 & 0 \\\cline{3-7}
& & SOLVED & \textbf{110} & 99 & 91  & 100 \\\hline
\multirow{2}*[-1.5ex]{20220314-Uncu} & \multirow{2}*[-1.5ex]{225} 
     & SAT & 69 & \textbf{70} & 62 & \textbf{70} \\\cline{3-7}
 & & UNSAT & \textbf{155} & 153 & 148 & 152 \\\cline{3-7}
& & SOLVED & \textbf{224} & 223 & 210 &  222 \\\hline
\multirow{2}*[-1.5ex]{hong} & \multirow{2}*[-1.5ex]{20} 
     & SAT & 0 & 0 & 0 & 0 \\\cline{3-7}
 & & UNSAT & 8 & \textbf{20} & \textbf{20} &  11 \\\cline{3-7}
& & SOLVED & 8 & \textbf{20} & \textbf{20} &  11 \\\hline
\multirow{2}*[-1.5ex]{hycomp} & \multirow{2}*[-1.5ex]{2752} 
     & SAT & \textbf{307} & 227 & 225 & 296 \\\cline{3-7}
 & & UNSAT & \textbf{2242} & 2201 & 2212 & 2155 \\\cline{3-7}
& & SOLVED & \textbf{2549} & 2428 & 2437 & 2451 \\\hline
\multirow{2}*[-1.5ex]{kissing} & \multirow{2}*[-1.5ex]{45} 
     & SAT & \textbf{33} & 10 & 17 & 14 \\\cline{3-7}
 & & UNSAT & 0 & 0 & 0 & 0 \\\cline{3-7}
& & SOLVED & \textbf{33} & 10 & 17 & 14 \\\hline
\multirow{2}*[-1.5ex]{LassoRanker} & \multirow{2}*[-1.5ex]{821} 
     & SAT & 167 & 122 & \textbf{305} & 303 \\\cline{3-7}
 & & UNSAT & 151 & 260 & \textbf{470} & 294 \\\cline{3-7}
& & SOLVED & 318 & 382 & \textbf{775} & 597 \\\hline
\multirow{2}*[-1.5ex]{meti-tarski} & \multirow{2}*[-1.5ex]{7006} 
     & SAT & \textbf{4391} & 4369 & 4343 & 4384 \\\cline{3-7}
 & & UNSAT & \textbf{2605} & 2588 & 2581 & 2591 \\\cline{3-7}
& & SOLVED & \textbf{6996} & 6957 & 6924 & 6975 \\\hline
\multirow{2}*[-1.5ex]{UltimateAutomizer} & \multirow{2}*[-1.5ex]{61} 
     & SAT & 35 & \textbf{39} & 35 & 37 \\\cline{3-7}
 & & UNSAT & 11 & 12 & 10 & \textbf{13} \\\cline{3-7}
& & SOLVED & 46 & \textbf{51} & 45 & 50 \\\hline
\multirow{2}*[-1.5ex]{zankl} & \multirow{2}*[-1.5ex]{166} 
     & SAT & \textbf{70} & 58 & 58 & 56 \\\cline{3-7}
 & & UNSAT & 28 & \textbf{32} & \textbf{32} & 30 \\\cline{3-7}
& & SOLVED & \textbf{98} & 90 & 90 & 86 \\\hline
\multirow{2}*[-1.5ex]{Total} & \multirow{2}*[-1.5ex]{12134} 
     & SAT & 5569 & 5372 & 5475 & \textbf{5619} \\\cline{3-7}
 & & UNSAT & 5379 & 5612 & \textbf{5809} & 5358 \\\cline{3-7}
& & SOLVED & 10948 & 10984 & \textbf{11284} & 10997 \\\hline
\end{tabular}
\vspace{2mm}
\caption{Summary of results for all instances in SMT-LIB (QF\_NRA).}
\label{tab:instances}
\end{table}

\begin{figure}
     \centering
     \includegraphics[width=0.15\textwidth]{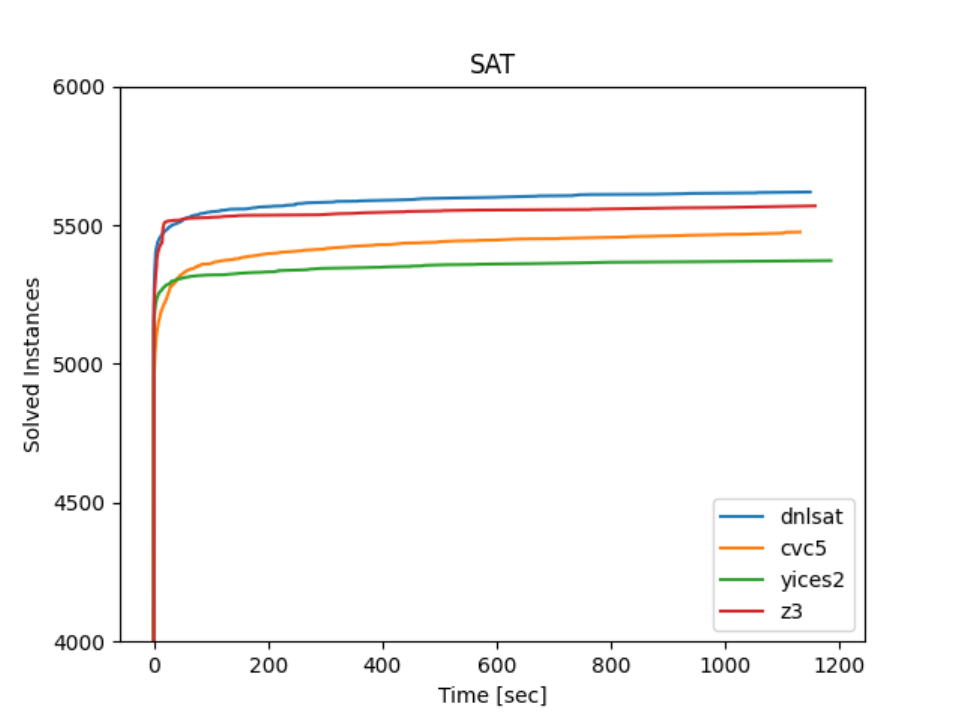}
     \includegraphics[width=0.15\textwidth]{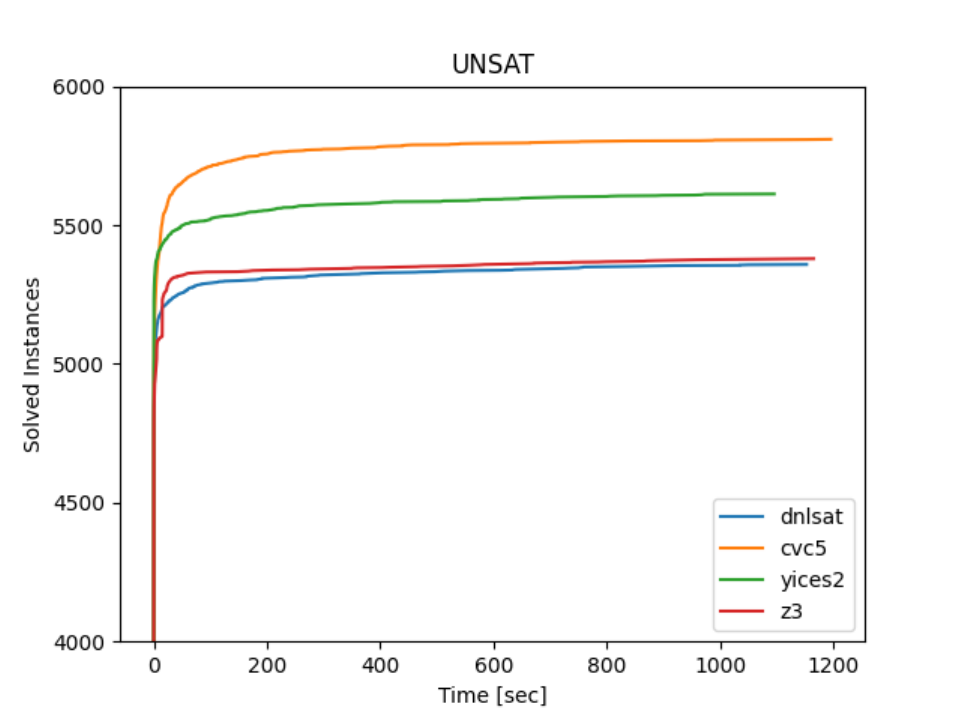}
     \includegraphics[width=0.15\textwidth]{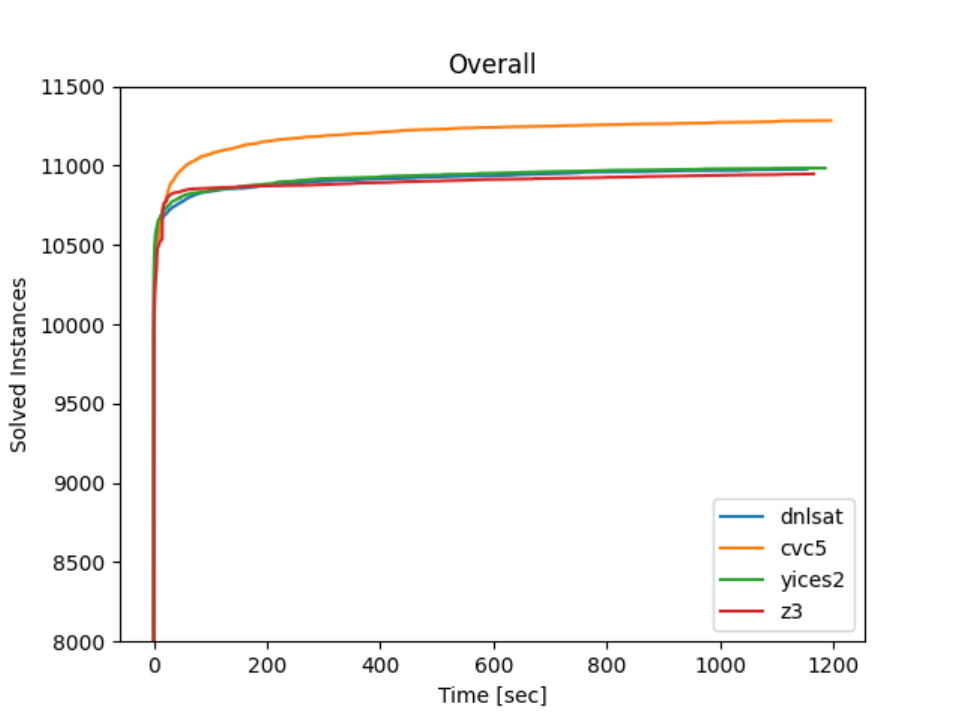}
    \caption{The number of solved instances by SMT solvers within different time limits.}
\label{fig:solvers}
\end{figure}

We also count the number of solved instances within different time limits by each solver in Figure ~\ref{fig:solvers}. It is shown that dnlsat solves almost the most satisfiable instances in any time range. Although dnlsat is not competitive at unsatisfiable instances, dnlsat solves almost the same number with Z3 solver. In conclusion, dnlsat increases performance on satisfiable instances, while does not bring much side effect on unsatisfiable instances.

\section{Conclusion}
We present a new SMT(NRA) solver called dnlsat, which implements an efficient dynamic variable ordering mechanism based on MCSAT. Several heuristics including projection order, lemma deletion and restart have also been incorporated into our decision procedure. Our experimental results demonstrate that dnlsat is competitive on most NRA instances, especially on satisfiable instances. 

\begin{acks}
The authors would like to thank the anonymous reviewers for
their comments and suggestions.
\end{acks}

\bibliographystyle{ACM-Reference-Format}
\bibliography{dnlsat}

\end{document}